\documentclass[twocolumn,showpacs,preprintnumbers,amsmath,amssymb]{revtex4}

\def\dd{{\rm d}}

\def\rv{{r_{_{\rm vir}}}}
\def\eb{{\epsilon_{_{\rm bar}}}}
\def\edm{{\epsilon_{_{\rm dm}}}}
\def\mdm{{M_{_{\rm dm}}}}
\def\cv{{c_{_{\rm v}}}}

\def\cvbar{{c_{_{\rm v,bar}}}}
\def\cvdm{{c_{_{\rm v,dm}}}}
\def\sb{{s_{_{\rm b}}}}
\def\sv{{\sigma_{_{\rm v}}}}
\def\mbar{{M_{_{\rm b}}}}
\def\nv{{N_{_{\rm vir}}}}
\def\nb{{N_{_{\rm bar}}}}
\def\ndm{{N_{_{\rm dm}}}}
\def\rt{{r_{_{\rm t}}}}
\def\rc{{r_{_{\rm c}}}}
\def\rs{{r_{_{\rm s}}}}
\def\rsc{{r_{_{\rm sc}}}}
\def\rdec{{r_{_{\rm dec}}}}
\def\rdecbar{{r_{_{\rm dec,bar}}}}
\def\rdecdm{{r_{_{\rm dec,dm}}}}

\def\lcdm{{\rm{ \Lambda CDM}}}
\def\kpc{{ \rm{ \,kpc} }}
\def\pc{{ \rm{ \,pc} }}
\def\gy{{ \rm{ \,Gyr} }}
\def\ms{{ \rm{ \,M_\odot} }}
\def\kms{{\rm{ \, km \, s^{-1}}}}
\def\la{\lower.5ex\hbox{$\; \buildrel < \over \sim \;$}}
\def\ga{\lower.5ex\hbox{$\; \buildrel > \over \sim \;$}}
\usepackage{epsfig}

\begin{document}

\title{Galaxy Satellites and the Weak Equivalence Principle}

\author{Jose Ariel Keselman}
\email{kari@tx.technion.ac.il}
\affiliation{Physics department, Technion, Haifa 
             32000, Israel}

\author{Adi Nusser}
\affiliation{Physics Department and the Asher Space Research Institute, 
             Technion, Haifa 32000, Israel}

\author{P.~J.~E. Peebles}  
\affiliation{Joseph Henry Laboratories, Princeton University, Princeton,
             NJ 08544, USA}

\begin{abstract}
Numerical simulations of the effect of a long-range scalar interaction
(LRSI) acting only on nonbaryonic dark matter, with strength comparable
to gravity, show patterns of disruption of satellites that can agree
with what is seen in the Milky Way. This includes the symmetric 
Sagittarius stellar stream. The exception presented here to the Kesden
and Kamionkowski demonstration that an LRSI tends to produce distinctly
asymmetric streams follows if the LRSI is strong enough to  separate
the stars from the dark matter before tidal disruption of the stellar
component, and if stars dominate the mass in the luminous part of the
satellite. It requires that the Sgr galaxy now contains little dark
matter, which may be consistent with the Sgr stellar velocity dispersion,
for in the simulation the dispersion at pericenter exceeds
virial. We present other examples of simulations in which a strong 
LRSI produces satellites with large mass-to-light ratio, as in Draco,
or free streams of stars, which might be compared to ``orphan'' streams.
\end{abstract}


\maketitle

\section{INTRODUCTION}
\label{sec:sec1}
Kesden and Kamionkowski (\cite{r1,r2}, hereafter KK06), in an independent
application of the point made by Frieman  and Gradwohl \cite{r3}, analyzed
an important astronomical test of the weak equivalence principle, 
that the acceleration of a freely moving test particle is independent
of its composition, applied to nonbaryonic dark matter (DM). They
pointed out that if the dark matter significantly violated the weak
equivalence principle then the disruption of a satellite falling close
to the host galaxy would tend to produce quite asymmetric stellar
streams, a result of the non-inertial  motion of the DM potential well
that is holding the stars. This effect is contrary to the symmetric
stellar streams of the Sagittarius (Sgr) satellite of the Milky 
Way (MW). The main purpose of this paper is to present a numerical
simulation that demonstrates an exception to the Kesden-Kamionkowski
effect. 

Our exception, symmetric stellar streams, requires two conditions.
First, the departure from the equivalence principle is large enough
that it separates the stars and dark matter in a satellite falling
toward its host galaxy before significant tidal disruption of the
luminous parts of the satellite. Second, the mass in the central region
of the satellite is dominated by stars. This second condition is
consistent with the observation that the stars in the Sgr 
galaxy and stream have heavy element abundance patterns \cite{r4} that are
considered to be characteristic of a more massive satellite such as the
Large Magellanic Cloud \cite{r5}, where 
the mass in the luminous central parts is dominated by stars \cite{r6}. We
show in the simulation in \S \ref{sec:sec3} how the stellar system once
separated from its dark matter halo suffers the usual pure
gravitational tidal disruption as it falls toward the denser parts of 
the host, and  can end up with a symmetric stellar galaxy and stream
with properties reasonably similar to the Sgr system.

A significant departure from the weak equivalence principle in the dark
sector would have other observable consequences.  A simulation in \S\ref{sec:sec4} is meant to
approximate the evolution of Draco, which seems to be dominated by DM,
perhaps because it is strongly enough bound and has remained far enough
from the galaxy to have avoided complete separation of the stars and
DM. Other simulations in this section show the development of a coreless stellar stream, which might be related to the Orphan streams in the MW \cite{r7}, and twin gravitationally bound cores, one dark and one
stellar. The latter has not been observed but might prove to be
interesting. 

Our computation follows previous studies (reviewed in \cite{r8}) in
expressing a departure from the weak equivalence principle as a fifth
force of interaction in the dark sector. We take the interaction to be
mediated by a scalar field that produces the fifth force, or
LRSI, 
\begin{equation}
{\bf F}=-\frac{\beta Gm^2}{r}e^{-r/\rs}\left( \frac{1}{r}+\frac{1}{\rs}
\right)\bf{\hat{r}},
\label{eq:eq1}
\end{equation}
between DM particles. Here $G$ is  Newton's gravitational constant, $m$
is the mass of a DM particle, $\bf{r}$ is the separation vector,
$\beta$ is a measure of the fifth force strength  relative to gravity,
and $\rs$ is a ``screening length". This fifth force applies only
between DM particles; it adds to the usual gravitational interaction
among DM and baryons. In the simulations presented here $r_s$ is much
larger than the size of the galaxy so the fifth force is in effect an
inverse square law. In the scalar field model for Eq.~(\ref{eq:eq1})
the screening length scales with redshift as $r_s\propto (1+z)^{-1}$ \cite{r8a}.
That is irrelevant for the effect on satellites of the MW but 
important in preventing any significant effect on the cosmological tests. 

KK06 showed that the ratio of stellar masses in leading and trailing
streams of a disrupting satellite can be quite sensitive to the value
of $\beta$, and they argued that the approximate symmetry of the
Sgr stellar streams implies $\beta<0.04$, in the
notation of Eq.~(\ref{eq:eq1}). Our alternative scenario for disruption
follows if $\beta$ is on the order of unity, meaning the DM fifth force
is about as strong as gravity on the scale of galaxies.

The next section outlines our numerical methods and parameter choices
for building satellites of a MW-like galaxy and simulating the effects
of gravity and an LRSI. The simulations use a version of the Gadget2
N-body code \cite{r9} modified to take account of the fifth force in
Eq.~(\ref{eq:eq1}). The numerical results for an Sgr-like satellite are
presented in \S~\ref{sec:sec3}. Because the picture requires that the
Sgr galaxy contains little DM we pay particular attention to the
departure of the stellar velocity dispersion from virial equilibrium as
the satellite moves around the galaxy. We argue in \S~\ref{sec:sec5}
that this effect can reconcile the measured star velocity dispersion in
the Sgr galaxy with a DM-free galaxy gravitationally bound by the mass
in its stars. In \S~\ref{sec:sec4} and~\ref{sec:sec5} we comment on other issues to be explored in more detail. 

\section{The numerical modeling}
\label{sec:sec2}
We seek reasonably realistic simulations of the effect of an LRSI on
observed satellites of the MW. In the simulations the galaxy is
represented as a static potential produced by a dark matter halo, a 
stellar disk, and  a stellar bulge. The behavior of the satellites 
is modeled by the N-body dynamics of two particle species,  
DM and stars, with the stellar component initially concentrated
well within the DM potential, as suggested by \cite{r10}, in initially 
spherical distributions. This description of
initial conditions and numerical methods is
supplemented by details in Appendix \ref{app:app1}. 

\subsection{Mass Distribution in the Host Galaxy}
The gravitational potential  of our model for the MW  is represented by the sum of 
three fixed components. The first is a spherical DM halo with potential 
\begin{equation}
\Phi_{\rm halo}(r) = v_{\rm halo}^2 \ln(r^2+b^2),
\label{eq:eq24}\vspace{3mm}
\end{equation}
where $r$ is the distance from the center of the galaxy, $v_{\rm halo
}=131.5\rm{\, km \, s^{-1}}$, and $b=12\kpc$. The second is a Miyamoto-Nagai
disk \cite{r11} with potential 
\begin{equation}
\Phi_{\rm disk}=-\frac{GM_{\rm disk}}{\sqrt{r_\perp^2+
(a+\sqrt{z^2+d^2})^2}},
\label{eq:eq25}
\end{equation}
where $z$ and $r_\perp$ are cylindrical coordinates and
$M_{\rm disk}=10^{11}\ms$, $a=6.5\kpc$, and $d=0.26\kpc$. The third is
a spherical Hernquist bulge \cite{r12} with potential 
\begin{equation}
\Phi_{\rm bulge}=-\frac{GM_{\rm bulge}}{r+c},
\label{eq:eq26}
\end{equation}
where $M_{{\rm bulge}}=3.4\times 10^{10} \ms$ and $c=0.7\kpc$.

The DM particles feel the additional force  given by the gradient of
$\Phi_{\rm scalar}=\beta\Phi_{\rm halo}$. We neglect the exponential
term in the scalar force in Eq.~(\ref{eq:eq1}) because the size of the 
galaxy is supposed to be much smaller than the screening length $\rs$.

\subsection{Construction of  the satellites} 
The initial DM and star distributions in a satellite are constructed to be spherically symmetric. The virial radius is 
$\rv=r_{_{200}}/\bigtriangleup$, where $r_{_{200}}$ is the distance within
which the mean mass density of DM plus stars is 200 times the present
critical cosmological density (for Hubble parameter
$H_0=$70 km~s$^{-1}$~Mpc$^{-1}$), and $\bigtriangleup$ is 
discussed in the next subsection. The
DM density profile at  $r\leq \rv$ is the Navarro,
Frenk \& White form \cite{r13, r14},  
\begin{equation}
\rho_{_{\rm DM}}(r) =
\frac{\rho_0}{{r/\rv}\left( 1+\cv r/\rv \right) ^2},
\label{eq:eq9}
\end{equation}
where $\cv$ is the concentration parameter and $\rho_0$ is a
normalization factor. At $r>\rv$ we choose an exponential DM density
profile, as in \cite{r15},
\begin{equation}
\rho_{_{\rm DM}}(r) = \frac{\rho_0}{\cv(1+\cv)^2}
\left( \frac{r}{\rv}\right)^{\alpha_{_{\rm DM}}}
\exp \left( -\frac{r-\rv}{\rdec}\right)\; .
\label{eq:eq10}
\end{equation}
The continuity of $\dd \rho_{_{\rm DM}}/\dd r$ at $r=\rv$ demands
\begin{equation}
\alpha_{_{\rm DM}} = \frac{-1-3\cv}{1+\cv}+\frac{\rv}{\rdec}.
\label{eq:eq11}
\end{equation}

The undisturbed inner stellar component follows the 
modified Hubble density profile \cite{r16},
\begin{equation}
\rho_{_{\rm b}}(r) = \rho_1
\left[ 1+  \frac{r^2}{\rc^2}\right]^{-2/3}, \hbox{ at } r<\rt, 
\label{eq:eq12}
\end{equation}
where $\rc$ is a core radius, $\rho_1$ is a normalization factor, and 
$\rt$ is a truncation radius of the stellar component. At $r>\rt$ the
 stellar density run is
 \begin{equation}
\rho_{_{\rm b}}(r) = \rho_1
\left( 1+\frac{{\rt}^2}{{\rc}^2}\right)^{-3/2}
\left( \frac{r}{\rt} 
\right)^{\alpha_{_{\rm b}}}
\exp \left( -\frac {r-\rt}{\rdec} \right)\; ,
\label{eq:eq13}
\end{equation}
where
\begin{equation}
\alpha_{_{\rm b}} =
\rt \left( \frac{1}{\rdec}-\frac{3\rt}{{\rc}^2+{\rt}^2} \right)\; ,
\label{eq:eq15}
\end{equation}
from continuity of $\dd \rho_{_{\rm b}}/\dd r$ at $r=\rt$.

We assign the satellite particles --- DM and stars --- statistically isotropic
velocity distributions that produce near dynamical equilibrium  prior to significant perturbation by the host
galaxy, using the variant of the method in \cite{r17} outlined in
Appendix \ref{app:app1}. Before placing the satellite in the
gravitational potential of the host galaxy we compute the motions of
the satellite particles for 20 gravitational dynamical times (as
measured at the scale radius $\rv/\cv$) using the Gadget2 N-body
code. The evolved density profiles are quite stable over this range of
time.

\subsection{Parameter choices} 
Throughout this study we set the ratio of the total DM to stellar mass
within a satellite to be close to the global cosmic value \cite{r18}. The stellar 
mass fraction may be smaller than that in a satellite, as it is in
larger galaxies (as reviewed in \cite{r16}), but that need not matter because excess DM in the
outermost parts of the satellite could be stripped away before much
happens to the stellar component. 

\begin{table}
\vspace{5mm}
\begin{center}
\begin{tabular*}{235pt}{@{\extracolsep{\fill}}|l|c|c|c|c|c|c|}
\hline
Simulation &S1 &S2 &S3 &S4 &S5 &S6\\
\hline \hline
1.  $\rc$           &$0.55$  &$0.55$  &$3.39$  &$2.61$  &$0.6$   &$0.6$  \\
2.  $\rsc$          &$3.7$   &$3.7$   &$3.39$  &$3.73$  &$1.73$  &$1.73$ \\
3.  $\rt$           &$1.67$  &$1.67$  &$10.17$ &$10.4$  &$1.81$  &$1.81$ \\
4.  $\rv$           &$18.65$ &$18.65$ &$17$    &$18.6$  &$8.65$  &$8.65$ \\
5.  $\rdecbar$      &$0.1$   &$0.1$   &$0.1$   &$0.1$   &$0.1$   &$0.1$  \\
6.  $\rdecdm$       &$0.1$   &$0.1$   &$0.1$   &$0.1$   &$0.1$   &$0.1$  \\
7.  $\cvbar$        &$3$     &$3$     &$3$     &$4$     &$3$     &$3$    \\
8. $\cvdm$          &$5$     &$5$     &$5$     &$5$     &$5$     &$5$    \\
9. $\sb$            &$0.15$  &$0.15$  &$1$     &$0.7$   &$0.35$  &$0.35$ \\
10. $\beta$         &$0$     &$1$     &$1$     &$1$     &$1$     &$2$    \\
11. $Y(t=0)$        &$80$    &$84.8$  &$170$   &$120$   &$140$   &$140$  \\
12. $v_{z}(t=0)$    &$80$    &$80$    &$140$   &$200$   &$280$   &$280$  \\
13. $\mdm(\infty)$  &$19$    &$19$    &$290$   &$19.3$  &$1.93$  &$1.93$ \\
14. $\mbar(\infty)$ &$4.15$  &$4.15$  &$62.3$  &$3.9$   &$0.42$  &$0.42$ \\
15. $\mdm(\rc)$     &$0.16$  &$0.16$  &$45$    &$1.9$   &$0.065$ &$0.065$\\
16. $\mdm(\rsc)$     &$3$     &$3$     &$45$    &$3$     &$0.31$  &$0.31$ \\
17. $\mdm(\rt)$     &$0.94$  &$0.94$  &$150$   &$9.3$   &$0.33$  &$0.33$ \\
18. $\mdm(\rv)$     &$15$    &$15$    &$227$   &$14.9$  &$1.5$   &$1.5$  \\
19. $\mbar(\rc)$    &$0.6$   &$0.6$   &$9$     &$0.48$  &$0.061$ &$0.061$\\
20. $\mbar(\rsc)$    &$4.15$  &$4.15$  &$9$     &$0.93$  &$0.29$  &$0.29$ \\
21. $\mbar(\rt)$    &$3$     &$3$     &$45$    &$3$     &$0.3$   &$0.3$  \\
22. $\mbar(\rv)$    &$4.15$  &$4.15$  &$60$    &$3.9$   &$0.42$  &$0.42$ \\
23. $\bigtriangleup$&$1$     &$1$     &$2.7$   &$1$     &$1$     &$1$    \\
24. $\nb$           &$5$     &$5$     &$1$     &$1$     &$1$     &$1$    \\
25. $\ndm$          &$1.4$   &$1.4$   &$1.4$   &$1.4$   &$1.4$   &$1.4$  \\
26. $\eb$           &$35$    &$35$    &$437$   &$437$   &$85$    &$85$   \\
27. $\edm$          &$650$   &$650$   &$650$   &$650$   &$331$   &$331$  \\
\hline
\end{tabular*}
\end{center}
\caption{Simulation parameters. See description in text.}
\label{tab:tab1}
\end{table}

We summarize in table \ref{tab:tab1} the parameters for the initial
structures of the six model satellites we have chosen 
to illustrate possibly interesting effects of an LRSI. Entries 1 to 6
are the scale lengths in $\kpc$ and 7 and 8 are the concentration 
parameters defining the structure of the model satellites, as defined
by Eqs.~(\ref{eq:eq9}) to~(\ref{eq:eq15}), prior to disturbance by
the host. 

The stellar component is embedded in a generally more
extended DM component. The parameter $\sb \equiv \rc / \rsc$ in
entry 9 is the ratio of the stellar to DM core radii.
According to \cite{r10}, the range of values of this parameter in MW 
satellites is quite large, ranging from perhaps $\sb\sim 0.05$ to
close to unity. 
The strength of the LRSI relative  to gravity is listed in entry 10.
The simulation S1 includes gravity alone ($\beta=0$).  

The host galaxy is centered on the origin of the coordinates, the disk is
centered on the $Z=0$ plane, and  the satellite moves in the YZ plane.
The initial satellite position is in the plane of the disk at the
distance (in kpc), from the center of the galaxy, given in entry 11. The
initial velocity in entry 12 (in km s$^{-1}$) is in the positive $Z$
direction.

Entries 13 to 22 give the initial stellar and DM masses, in units
of $10^8\ms$, within various of the radii listed in entries 1 to 4. 
The overdensity used to set the virial radius is 
$200 \times\bigtriangleup^3$, where $\bigtriangleup$ is listed in 
entry 23. The exception to 
$\bigtriangleup=1$ is S3, where $\bigtriangleup=2.7$. This
satellite is supposed to correspond to a high concentration ($\cv=13.5$)
DM halo. However, the number of particles needed for proper numerical
modeling is  proportional to the fourth power of the concentration
(see Appendix \ref{app:app1}). This becomes numerically expensive, 
and, as a compromise, we work with a larger $\bigtriangleup$ to
imitate the desired larger concentration.

The  parameters $\nb$ and $\ndm$ are the numbers of particles, in units of
$10^4$, representing the mass distributions in stars and
DM. The parameters $\eb$ and $\edm$ are the lengths in parsecs  of the
smoothing of the inverse square law for DM and stars in the numerical
simulations, as defined in the Gadget2 code.

Six simulations for the initial conditions in
Table \ref{tab:tab1} have been run with Gadget2. Only S1 does
not include LRSI; it serves as a benchmark for comparison to the
standard picture. Simulation S2 has  initial conditions similar to  S1
but is run with LRSI at the strength of gravity ($\beta=1$).  It is
intended to produce the general properties of the Sgr galaxy and stream. The satellite S3 is constructed with $\sb=1$, and has a DM-dominated core. It becomes almost completely dominated by DM as a result of  an LRSI. It could be an analogue of 
Draco, though it is not intended to be a close match to any of its
properties other than the high $M/L$ and the apparent absence of
tidal streams. In S4 the stellar core is destroyed while
the DM core survives, leaving a smooth stellar stream. Simulations
S5 and S6 are attempts to illustrate the formation of
twin cores, DM and stellar. They share the same initial conditions
but differ by the LRSI strength $\beta$.
For the Gadget2 code parameters, we chose conservative values,
resulting in conservation of at least $99.5\%$ of the total energy for all simulations.

\begin{figure}
\centerline{\epsfig{figure=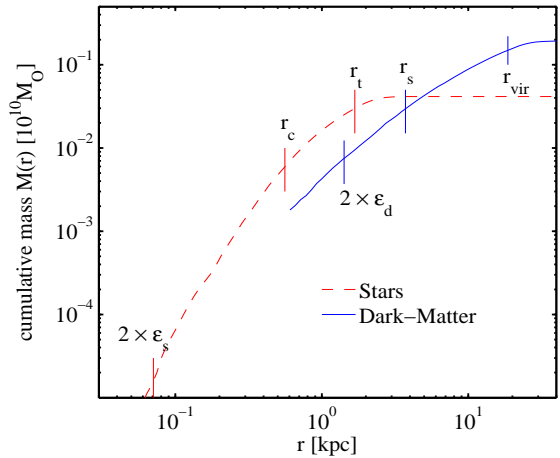, width=230pt, height=184pt}}
\caption{The initial stellar and DM masses as functions of
distance ${\rm r}$ from the center of the satellites S1 and S2.}
\label{fig:fig11}
\end{figure}

\section{Modeling the Sagittarius system}
\label{sec:sec3}

We describe here the evolution of model satellites S1 and S2
that are 
meant to simulate the formation of the Sgr galaxy and stream without
and with the effect of a strong departure from the weak 
equivalence principle represented by an LRSI in the dark sector.
We compare the models to observations in \S \ref{sec:sec5}. 

Initially the cores of S1 and  S2 are 
dominated by the stellar mass
out to $4\kpc$ radius, as illustrated by the DM and stellar mass
density runs in Fig.~\ref{fig:fig11}. As we have noted, and will
discuss in more detail in \S\ref{sec:sec5}, this is consistent with the
fact that the Sgr stars have been compared to the population in the
Large Magellanic Cloud (the LMC; e.g. \cite{r5}), in which the evidence is
that the mass in the central region is dominated by stars \cite{r6}. 

\begin{figure*}
\centerline{\epsfig{figure=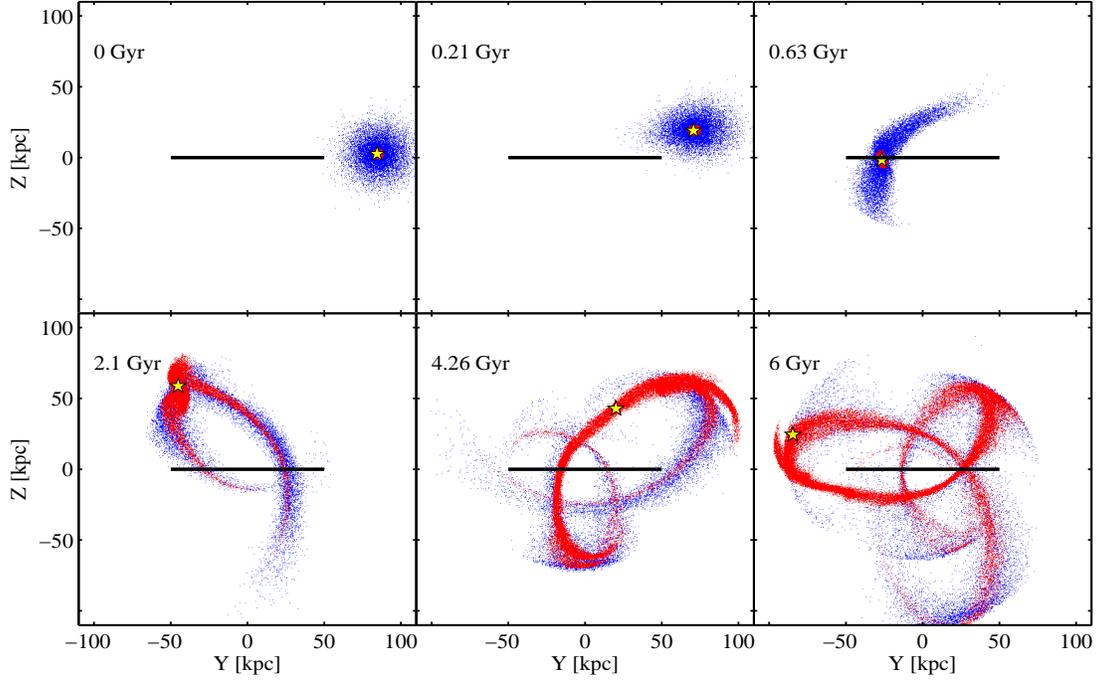, width=495pt, height=297pt}}
\caption{Snapshots in the YZ plane of the evolution of the S1 satellite
(without an LRSI). The stellar and DM 
particles are plotted as red open circles and blue dots, respectively.
For clarity, only a subset of the stellar particles are plotted.
The yellow
 star indicates the location of the most bound stellar particle.}
\label{fig:fig6}
\end{figure*}

\begin{figure*}
\centerline{\epsfig{figure=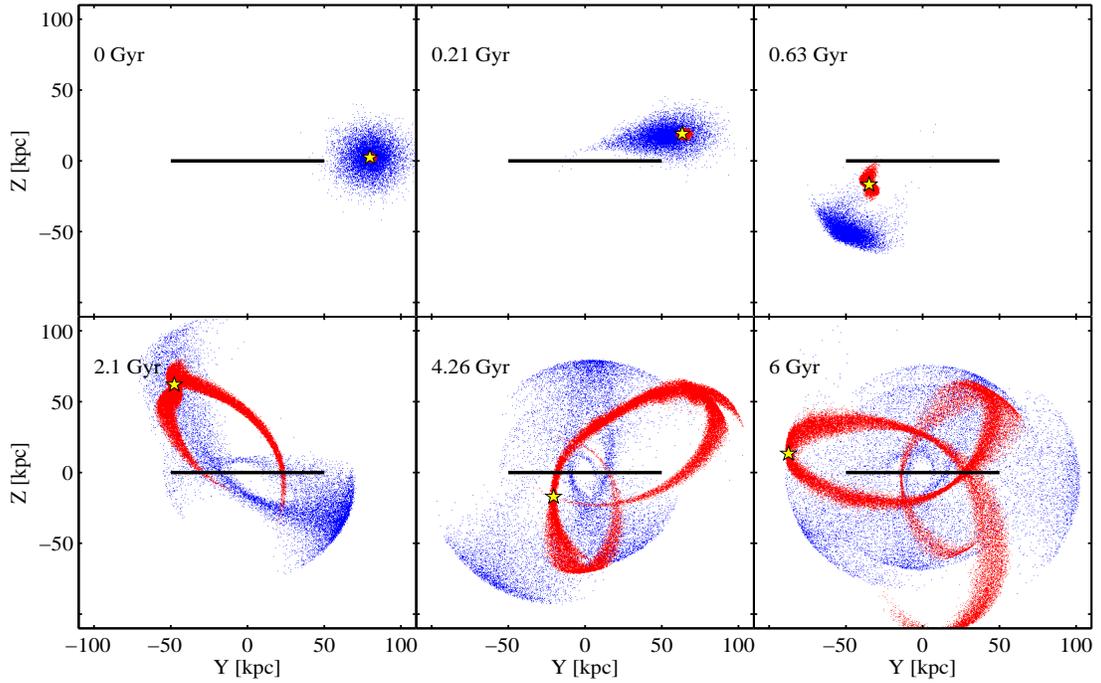, width=495pt, height=297pt}}
\caption{The same as Fig.~\ref{fig:fig6} but for S2 with an LRSI with
$\beta=1$.}
\label{fig:fig8}
\end{figure*}

To get  similar orbits we have assigned the same initial velocity but 
slightly different initial positions for S1 and S2  in the host galaxy (entries 11 and 12 in table~\ref{tab:tab1}). Both simulations are run for $6 \gy$. 

Evolutions of the
particle distributions projected in the $YZ$ plane perpendicular to
the plane of the galaxy are shown in Figs.~\ref{fig:fig6} and \ref{fig:fig8}. The
stellar core in S1 remains embedded in its  DM halo as the satellite
develops quite symmetric tidal streams. Early on the streams contain
mainly DM, but after the first pericenter passage stars become
prominent in the streams. 
The evolution of S2 is quite different, but it
ends up with similar symmetric stellar streams. In a frame of
reference moving with the DM core, the stellar core experiences an
effective force that works against the gravitational pull of the DM
core.  At $\beta=1$ and at the initial position of S2 this effective
force overcomes  the gravitational pull of the DM core. The result is
a nearly complete segregation of the stars and DM well before the
first pericentric passage. The separated stellar core is then free to
develop symmetric stellar tidal streams under purely gravitational
dynamics. Meanwhile the DM core is completely disrupted.

The behavior of S2 is very different from the KK06 simulations.
They focus on cases where the strength of the LRSI is weak enough that
there is not segregation of the stellar and DM components prior to
significant tidal disruption of the stellar system. In the KK06
simulations the effective force on the stars in the DM halo rest frame
tends to pull stars  out of one side of the halo, producing quite
asymmetric stellar streams.  

\begin{figure*}
\centerline{\epsfig{figure=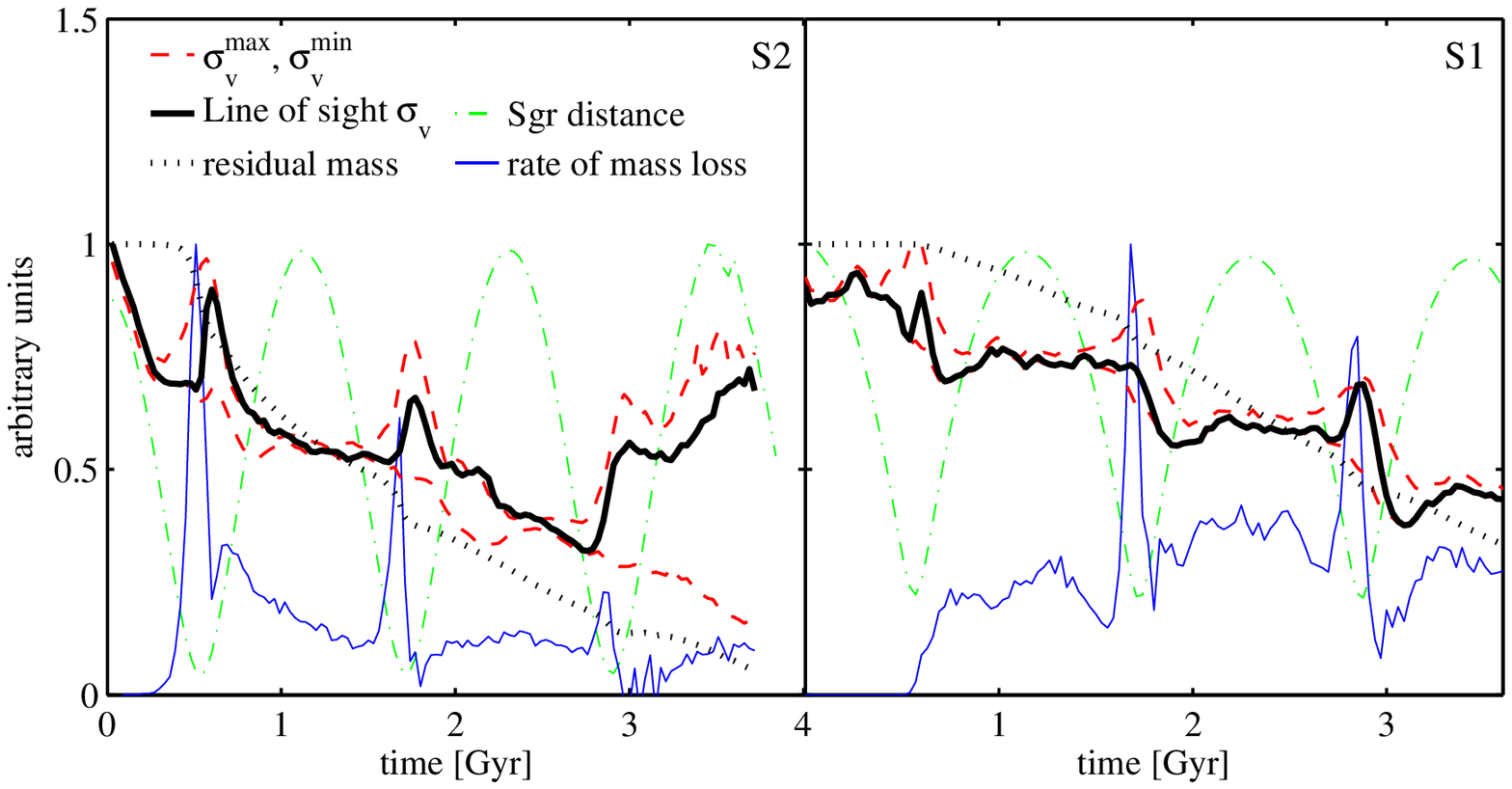, width=380pt, height=184pt}}
\caption{Measures of evolution of the S1 (panel to
the right) and S2 (to the left) simulations.}
\label{fig:fig1}
\end{figure*}

Fig.~\ref{fig:fig1} shows measures of the evolution of S1 and S2. 
The vertical scales of each curve are arbitrarily normalized to 
fit the figure, independently in each panel.
The dot-dashed curves are the satellite distance from the
center of the host, showing pericenter passages. The solid blue curves
show the rate of loss of the satellite stellar mass, and the
dotted curves show the residual stellar mass. Here,
the satellite is defined as the set of self-bound particles, which for
S1 is calculated with both DM and stellar particles, and for S2 with
stellar particles alone. As expected, the  mass loss rate in both
simulations is highest during pericenter passages where the tidal
force is strongest. The figure also shows
one-dimensional stellar velocity dispersions measured within
$0.7\kpc$  projected distance from the center of the satellite. 
The line-of-sight
velocity dispersion for an observer at ($0,-8.5\kpc,0$) relative to
the center of the host galaxy is plotted as the thick black lines. The
red dashed lines show the dispersions in the directions giving the
maximum and minimum values. In these simulations
the line-of-sight direction is 
close to the direction of maximum velocity dispersion at pericenter
passages.

The satellite is constructed to be in dynamical equilibrium with an 
isotropic velocity distribution prior to disturbance by the host.
The anisotropy that develops during pericenter 
passages, and the increase in the line-of-sight velocity dispersion,
are combined results of the contamination of escaping particles 
and the transient departure from dynamic equilibrium, including
the distortion of the satellite shape caused by the strong tidal
force. The deviations from an isotropic stellar velocity distribution
become larger in S2 once the stars are separated from the DM, 
the residual DM in S1 keeping the stars more tightly bound.

\begin{figure*}
\centerline{\epsfig{figure=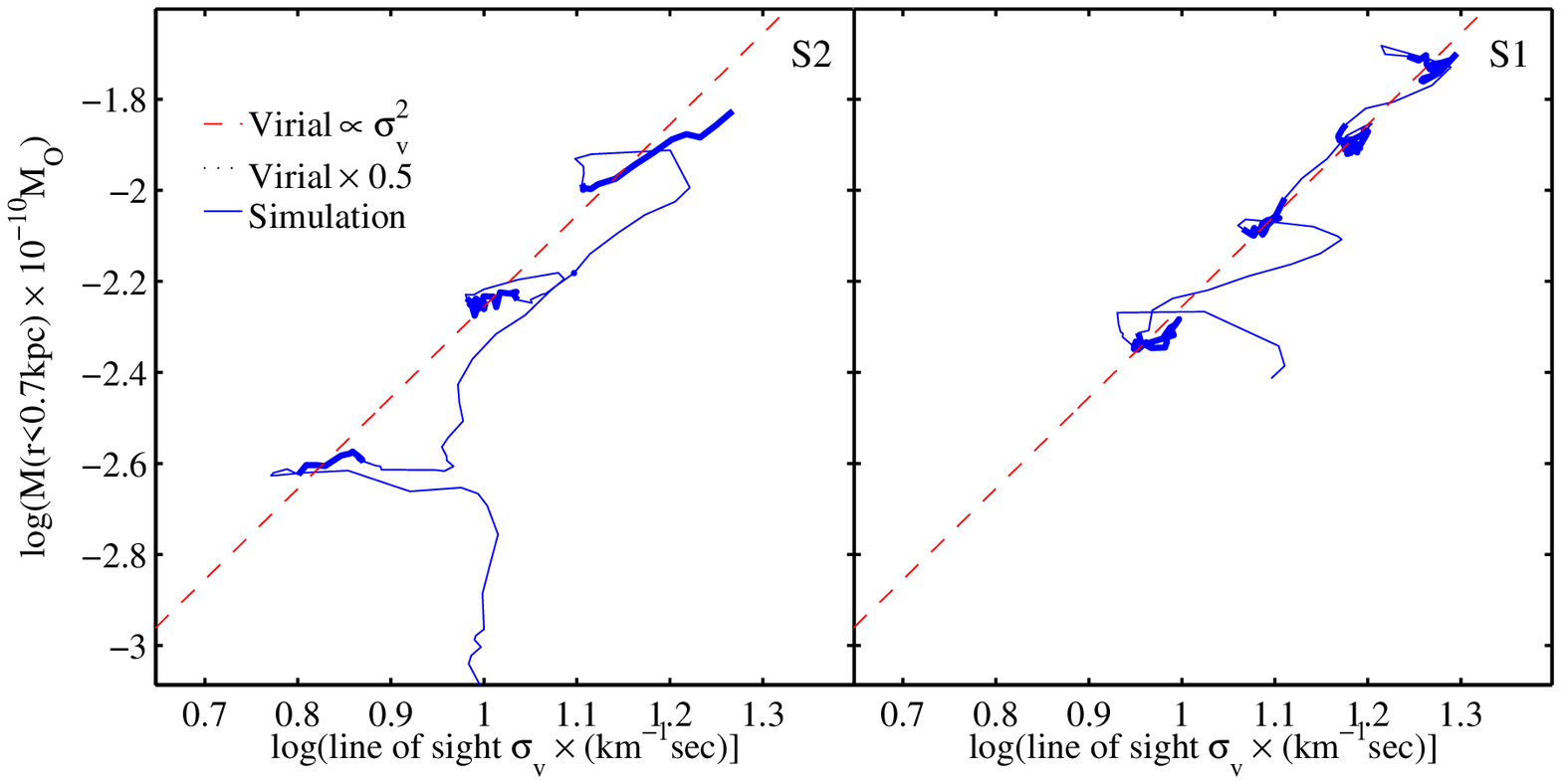, width=380pt, height=184pt}}
\caption{Evolution of the satellite bound stellar mass and stellar 
line-of-sight velocity dispersion follows the blue lines commencing
near the top right corner of each panel and ending toward the lower left. At the thick portions
the velocity anisotropy is less than 15\%. The dashed line varies as
$\sv^2$, as expected for dynamical equilibrium
in a satellite with fixed size.}
\label{fig:fig4}
\end{figure*}

Fig.~\ref{fig:fig4} illustrates how transient disturbances to the
stellar velocity distribution can affect dynamical mass estimates of the satellites S1 and S2. The 
blue curves show the evolution of the stellar mass and the stellar line-of-sight velocity dispersion $\sv$, both computed within projected radius $0.7\kpc$ perpendicular to the line of sight.
The thick portions of the 
curves correspond to times when $\sv^{max}-\sv^{min}<0.15\sv^{mean}$,  where $\sigma^{mean}$ is the RMS value averaged over all directions. 
The satellites start at the top-right corners and  end toward the lower-left after $4.17\gy$  for S1 and  $3.3\gy$ for S2.
 The dashed red lines show how the velocity dispersion $\sv$ would vary with the mass, $M(<r)\propto \sv^2$, at dynamical equilibrium and fixed radius $r$. The blue curves pass close to this condition when the velocity anisotropy is small.  At the thin portions of the line the satellite is closest to the host galaxy and most disturbed. Here the mass is smaller than expected from the 
velocity dispersion $\sv$ under the assumption of dynamical
equilibrium,  by a factor as large as two and even larger for S2 towards the final time.  Earlier studies of 
the possible  importance of tidally-driven departures 
from dynamical equilibrium differ, but some indicate
that velocity dispersions of long-lived dwarf satellites can 
be interpreted as the effect of a departure from equilibrium 
rather than the presence of dark matter \cite{r19,r20,r21,r22}. 

\begin{figure}
\centerline{\epsfig{figure=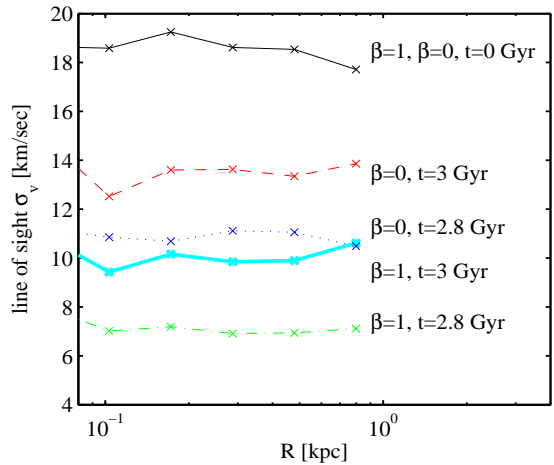, width=230pt, height=184pt}}
\caption{Line-of-sight stellar velocity dispersions as functions of
projected distance from the satellite center are shown for S1 ($\beta=0$) and S2
($\beta=1$) at various  times indicated in the figure.}
\label{fig:fig13}
\end{figure}

Fig.~\ref{fig:fig13} shows that the stellar velocity dispersions in
the cores of S1 and S2 are close to constant across the face of the
galaxy.
The curves at $t=3\gy$ are higher than at $t=2.8\gy$ in 
both S1 and S2, because the former coincides with a
pericentric passage in both simulations. 

\begin{figure*}
\centerline{\epsfig{figure=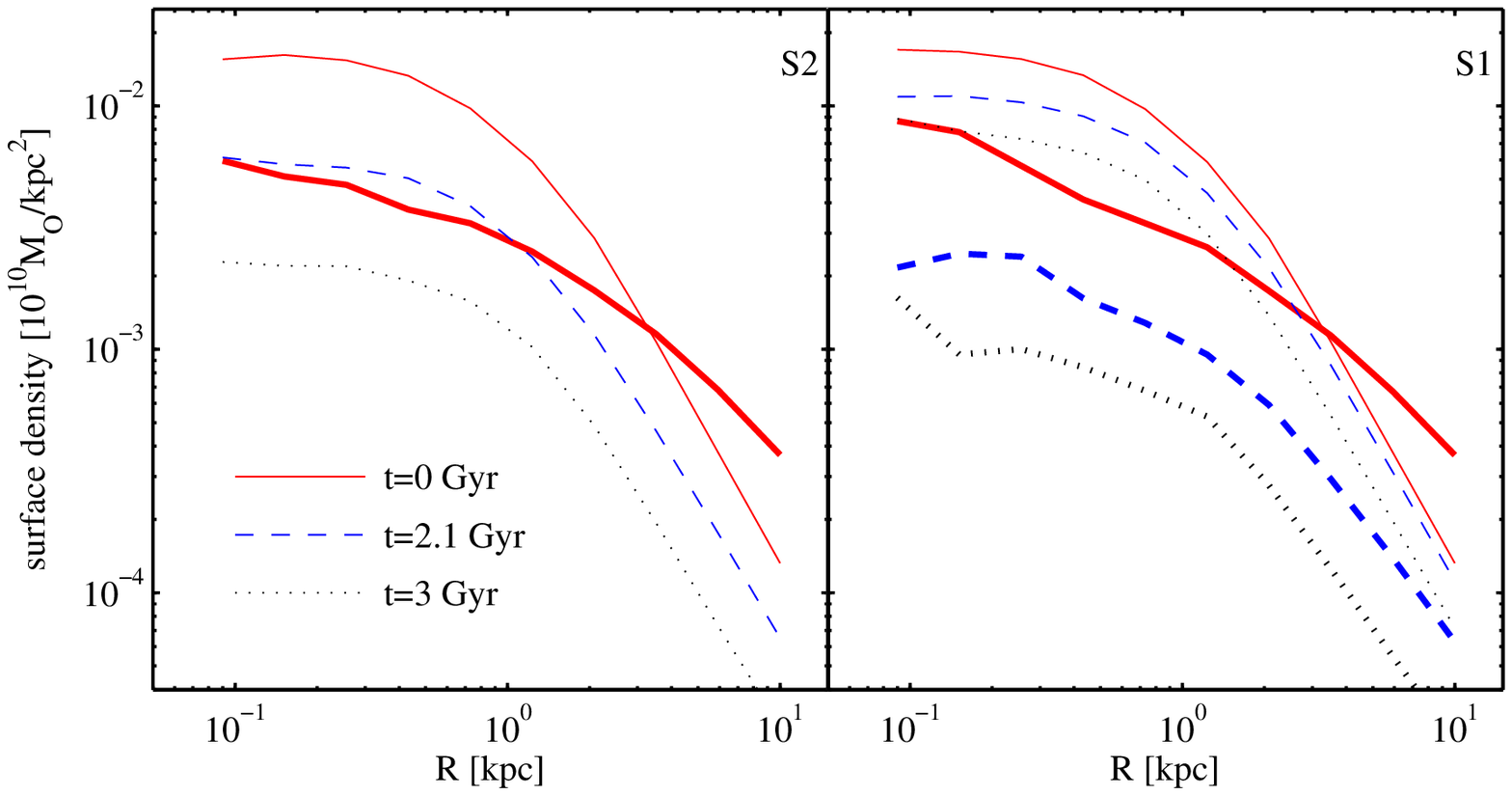, width=380pt, height=184pt}}
\caption{The mean stellar surface density profiles at different times
for S1 (right panel) and S2 (to the left). The DM profiles for S1 are
plotted as the thicker lines.}
\label{fig:fig5}
\end{figure*}

The evolution of
the mean surface density profile as a function of projected 
distance $R$ from the center of the satellite is shown in 
Fig.~\ref{fig:fig5}. The stellar component plotted as the thin lines
preserves a flat density in the inner regions, in agreement with \cite{r46}

The parameter KK06 use as a measure of the degree of symmetry of the
tidal stream is the ratio of the number of leading to trailing stars. 
KK06 find that in all their simulations without an LRSI this ratio is
never lower than $0.5$, while in the simulations with $0.04 \le \beta
\le 0.09$ it never exceeds $0.2$. Indeed, this supports the conclusion
that a  weak LRSI is seriously constrained by the observations of Sgr.
In the strong LRSI case we are considering, there is some
asymmetry in the early stages of evolution, which begins to decay right after 
the  DM leaves the
stellar core. At $t=1.7\gy$ the ratio of the number of leading to trailing stars, excluding particles that are closer than $4\kpc$ to the center, is $1.09$ and $0.817$ for S1 and S2 respectively. At
$t=2.85\gy$ the ratios are $0.96$ and $0.88$. 

We discuss how these models for the Sgr galaxy and stream compare 
to the observations in \S \ref{sec:sec5}, after presenting simulations
of a few other situations of possible astronomical interest.   

\section{Other manifestations of an LRSI}
\label{sec:sec4}\label{sub:sub2}

The mass-to-light ratio of the Draco satellite, at distance $D\approx 80 \;
\kpc$ from the MW,  is estimated to be $M/L > 100$ Solar units
\cite{r23}. This large value is not reasonably explained by deviations from
equilibrium \cite{r24} such as those occurring near pericenter passages
discussed in the previous section; it more likely means the mass
is dominated by DM throughout this satellite.
In the standard scenario, tidal mass loss may result
in an increased $M/L$ ratio \cite{r46}, however this effect
is not strong enough to explain highly DM dominated
satellites as Draco.
LRSI offers a mechanism
that can work either separately or in tandem with the usual postulate
that photoionization heating by the UV background has  
suppressed star formation \cite{r25,r26,r27}, provided
the LRSI is strong enough to have stripped away most of the stars
but weak enough to leave the most tightly bound stars in a DM halo that remains bound.  

\begin{figure*}
\centerline{\epsfig{figure=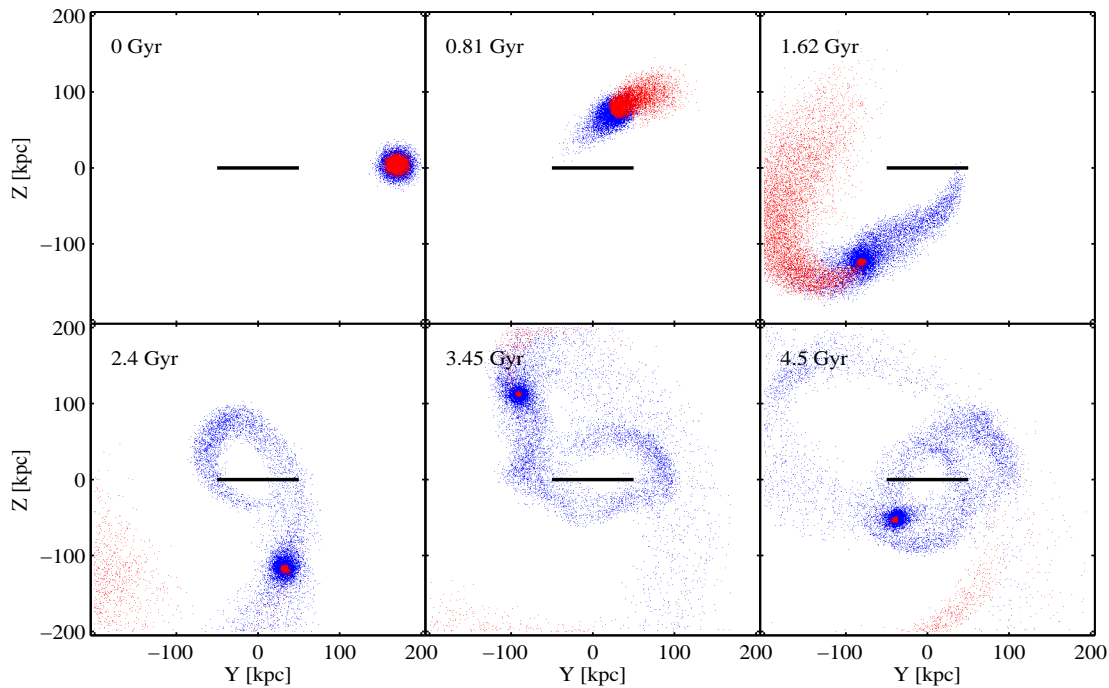, width=495pt, height=297pt}}
\caption{Illustration S3 of the development of a satellite that retains a small but nonzero fraction of its stars. The orientation is
the same as in Figs.~\ref{fig:fig6} and \ref{fig:fig8}. Red open circles are
the stellar component, blue dots the DM.}
\label{fig:fig14}
\end{figure*}

Our simulation S3, with $\beta =1$, illustrates a case of significant but not 
complete loss of stars in the LRSI picture. The 
stellar and DM core radii initially are the same 
($\sb=1$ in table \ref{tab:tab1}), and the 
central region is dominated by DM (at close to the cosmic
baryonic to DM mass ratio). One sees in Fig.~\ref{fig:fig14}
that the LRSI pulls most of the stellar 
particles out of the DM halo early in the evolution, in the direction opposite to the motion. That leaves a bound DM halo with a tighter concentration
of stars, in which the ratio of bound DM mass to stellar mass is 16 times larger
than  the initial value. 

\begin{figure*}
\centerline{\epsfig{figure=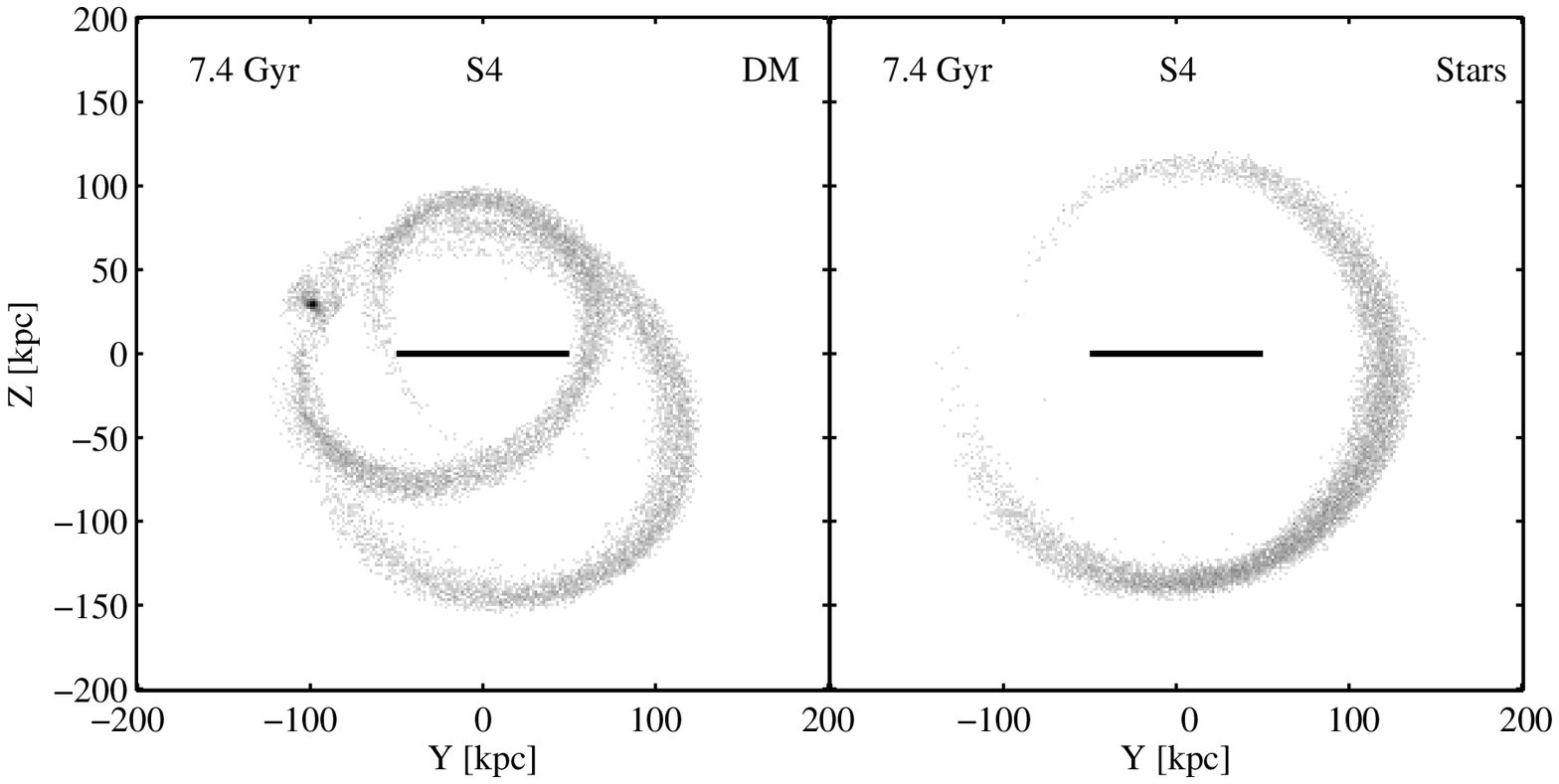, width=380pt, height=184pt}}
\caption{Mass distributions near the end of simulation S4.}
\label{fig:fig15}
\end{figure*}

Simulation S4 illustrates the
development of the coreless stellar stream in 
Fig.~\ref{fig:fig15} that might be compared to the 
orphan streams in the MW. In this simulation the stellar core is not 
gravitationally self-bound (the satellite core mass is not dominated by stars) so it is disrupted by the
segregation driven by the strong LRSI. Even after $7.4\gy$ the
stars remain in a smooth stream, little disturbed by 
the small mass left in the DM core in this simulation. 

\begin{figure*}
\centerline{\epsfig{figure=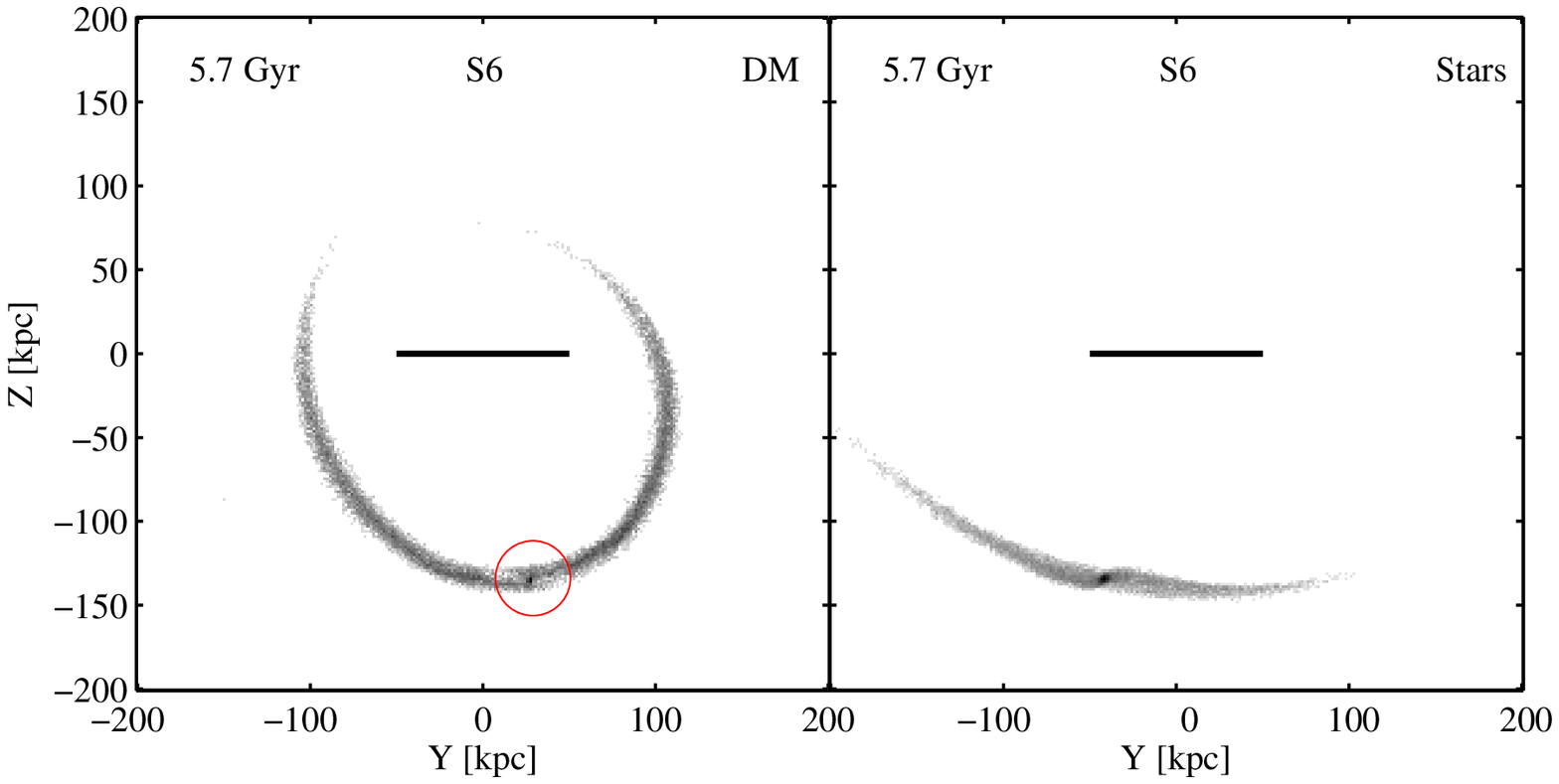, width=380pt, height=184pt}}
\caption{Simulation S6. Circled in red is the
remnant DM core.}
\label{fig:fig16}
\end{figure*}

Simulations S5 and S6 are tuned to explore the possibility that both
the  DM and stellar cores survive separation by an LRSI, ending up as
separate self-bound objects. Since a DM core is bound by the combined
effect of gravity and the LRSI its binding energy is enhanced by the
factor $1+\beta$ relative to a stellar system with the same mass
distribution. This means that, once we arrange for  a self-bound
remnant stellar core, a DM core will survive if  $\beta$ is large
enough. We did not find bound twin cores in S5 with $\beta=1$.
Fig.~\ref{fig:fig16}, for $\beta=2$ (S6), shows twin cores, but the remnant bound DM halo mass is  only 1\% the mass of the original satellite.

\section{DISCUSSION}
\label{sec:sec5}
The central point of this paper is that a strong departure from the
weak equivalence principle for nonbaryonic dark matter, with
$\beta\simeq 1$ in the LRSI model in Eq.~(\ref{eq:eq1}), can produce a
system that resembles the Sgr galaxy and stream from a progenitor that
seems reasonable from the point of view of the astronomy.  We have
also taken note of other modes of satellite disruption that might be
observationally interesting. 

It is important for our picture of the origin of the Sgr system that
the mass distribution in the central part of the Sgr progenitor is
dominated by stars. The thought that this might be so is motivated by
the observation that stellar element abundances in the Sgr galaxy and 
stream have been compared to LMC stars (e.g. \cite{r5}), and the evidence 
that  the mass in the luminous central parts of the LMC is dominated 
by stars \cite{r6}. 

We have aimed for a progenitor in our simulation that has structure
similar to the LMC (though a rough similarity is likely 
all that could be meaningful). Our S1 and S2 cases each
have initial total mass $\approx 2.2 \times 10^9 \ms$. The virial
radius is slightly less than 20 kpc, and stars dominate the mass
within 4 kpc radius.  Kim et al. \cite{r6} find that the LMC mass is
$\approx 3.5\times 10^9\ms$ within $4\kpc$ radius, and the total disk
mass is $2.5\times 10^9\ms$ within $7.3\kpc$ radius. By these
measures  the initial structures of S1 and S2 are reasonably 
similar to the LMC.

In S2, with the strong LRSI, the bound  mass in stars at
the third pericenter passage at $3\gy$ is $5.2\times 10^7\ms$. 
At stellar mass-to-light ratio $M/L = 2$ Solar units the luminosity
would be $L\sim 1\times 10^8L_\odot$. This is fairly close to the
measured luminosity of the Sgr galaxy, in the range
$(2-5.8)\times 10^7 L_\odot$ \cite{r29, r30}.

In our model the Sgr galaxy no longer contains dark matter. Estimates of
$M/L$ for this galaxy are larger than expected for stars; thus
from a recent program of observations \cite{r31} finds $M/L=17.5$. This
assumes dynamical equilibrium, however, and we have
illustrated in Figs.~\ref{fig:fig1} and \ref{fig:fig4} how the galaxy
may be disturbed from equilibrium by its host. In S2 the 
satellite at $3\gy$ is at pericenter $18\kpc$ from
the center of the host galaxy, similar to the present distance of the
Sgr galaxy, $16\kpc$ \cite{r32}. Also, although our 
host galaxy is rigid its mass distribution is designed to resemble that of
the MW. At this third pericenter passage the line-of-sight 
velocity dispersion within the
innermost $0.7\kpc$ of the simulated satellite is close to constant
across the core at $\sigma=9.8\kms$ (Fig.~\ref{fig:fig13}). This is in
line with the finding of \cite{r31} that the velocity dispersion in the Sgr
galaxy  is close to constant across the inner $0.7\kpc$ from the
center defined by M54 at $\sigma=9.6\kms$.  

The departure from dynamical equilibrium at pericenter, and
the large apparent value of the mass, also occurs in conventional
gravity without the LRSI, as in S1, and in the 
simulations by \cite{r33, r34}. The effect is larger in S2 than S1, however,
because of the greater fragility of the stellar core after
the early loss of the DM.

Our conclusion from these comparisons of S2 to the observations is
that it is possible to obtain a reasonably realistic picture of the 
origin of the Sgr system, including a DM-free galaxy and a symmetric stream, under the effect of a strong departure from the weak equivalence principle in
the dark matter. It would be interesting to study the effects of LRSI
on the features of the tidal streams, as done in \cite{r47}.

Our simulations suggest alternative interpretations of other observations.
The example in Fig.~\ref{fig:fig14}, a satellite that loses most but
not all of its stars, might be compared to Draco. A rough analytic
argument that the Draco DM halo 
can retain its most tightly bound stars against the effect of a strong
LRSI goes follows.  If the MW rotation curve for baryons is close to
flat at $v_c=220$ km~s$^{-1}$ at distance $D$ then a star in the DM
halo rest frame experiences acceleration $\beta v_c^2/D$. The
gravitational acceleration produced by the mass in the satellite at
distance $r$ from its center is maximum at
$GM(<r)/r^2\simeq 2\sigma^2/r_{_{c}}$ at the core radius $r_{_{c}}$.
According to \cite{r23} the stellar core radius of Draco  is $160\pc$. 
These numbers indicate that the one-dimensional stellar velocity dispersion
must be $\sigma=7\beta^{1/2}$ km~s$^{-1}$ to keep stars in the core of
Draco at its present distance $D$. Since the measured dispersion is in
the range $\sigma = 8.5-10.7 \; \kms$ \cite{r23}, we conclude that a strong
LRSI, $\beta\sim 1$, allows Draco to contain stars  as long as it has
not ventured much closer to the MW than its present distance. 

In the numerical simulation in Fig.~\ref{fig:fig14} there are remnant
bound stars, the ratio of bound DM to stellar mass is
an order of magnitude larger than the
original value with the DM remaining dominant at all radii, and the  
stars that
have escaped are broadly scattered. This is
in line with the properties of Draco.  Whether more detailed measures
of the simulated remnant satellite can be consistent with what is
observed in Draco is a subject for separate study.

Another example for closer study is the difference between the observed numbers of satellites and the far greater numbers of DM halos predicted by the $\lcdm$ cosmology \cite{r35, r36, r37}. This usually is taken to mean that many DM halos are not visible because they were unable to accrete or retain
photoionization-heated plasma \cite{r25, r26, r27}. The 
possibility offered by an LRSI is 
that many  halos are not visible because stars have been removed
from their DM halo, perhaps largely, as may have happened to Draco, or entirely. 

Depending on the structure of the
progenitor, stars removed from the DM may be broadly dispersed or 
may end up in a coreless stream, as in Fig.~\ref{fig:fig15}. The latter
offers an interpretation the orphan streams
in the MW (e.g. \cite{r7}). Orphan streams can
form by tidal disruption in standard gravity acting on
satellites in pericentric passages close enough or repeated often
enough to make it difficult to identify the remnant satellite or totally disrupt
it. In this scenario the formation time of coreless streams could be
quite long \cite{r22}. In the LRSI scenario demonstrated in Fig.~\ref{fig:fig15}
coreless streams may form rapidly, and could be found  in orbits with
large pericentric distances. This behavior might be detectable, and is
a signature of a strong LRSI.

Some stars separated from the DM may end up in gravitationally
bound DM-free dwarf galaxies. There are discussions of dwarfs with
mass-to-light ratios that might be considered characteristic of a
normal stellar population \cite{r28}. DM-free dwarfs  would be an interesting challenge to the standard theory. If the stars in the progenitor were much more strongly concentrated than the DM then a normal gravitational tide
could strip away most of the DM, leaving a bound system
dominated  by stars, but it would be a delicate operation. An
LRSI offers an alternative that operates in a straightforward way. 

A recent study \cite{r44} challenges the model presented here.
The main concern is that the satellite S2 is tightly constrained in the initial distance in which
it may first enter the MW, since it is too light for
dynamical friction to modify its orbit into the observable one.
It is important to notice that in the
standard model (without LRSI), Sgr could have many different  possible 
orbital histories
which lead to its currently observed setup. According to \cite{r43}, in one family of solutions the Sgr may have 
entered the MW at a galactocentric distance
close to its current apocentre of $60 \kpc$, with a mass of $10^9 \ms$. In this model, the Sgr 
is indeed too
light for gravitational friction to have any significant effect on its
orbit. This model is consistent with our satellite model S2. In another distinct 
family of solutions, the Sgr
could enter the MW at a distance of more than $200 \kpc$ with a total mass  of about
$10^{11} \ms$ which is 100 times more massive than S2. 
In this scenario, dynamical friction plays a key role in
bringing Sgr to its current orbit, while along the way it looses more
than 99\% of its mass. This scenario is consistent with \cite{r45}.

\begin{figure*}
\centerline{\epsfig{figure=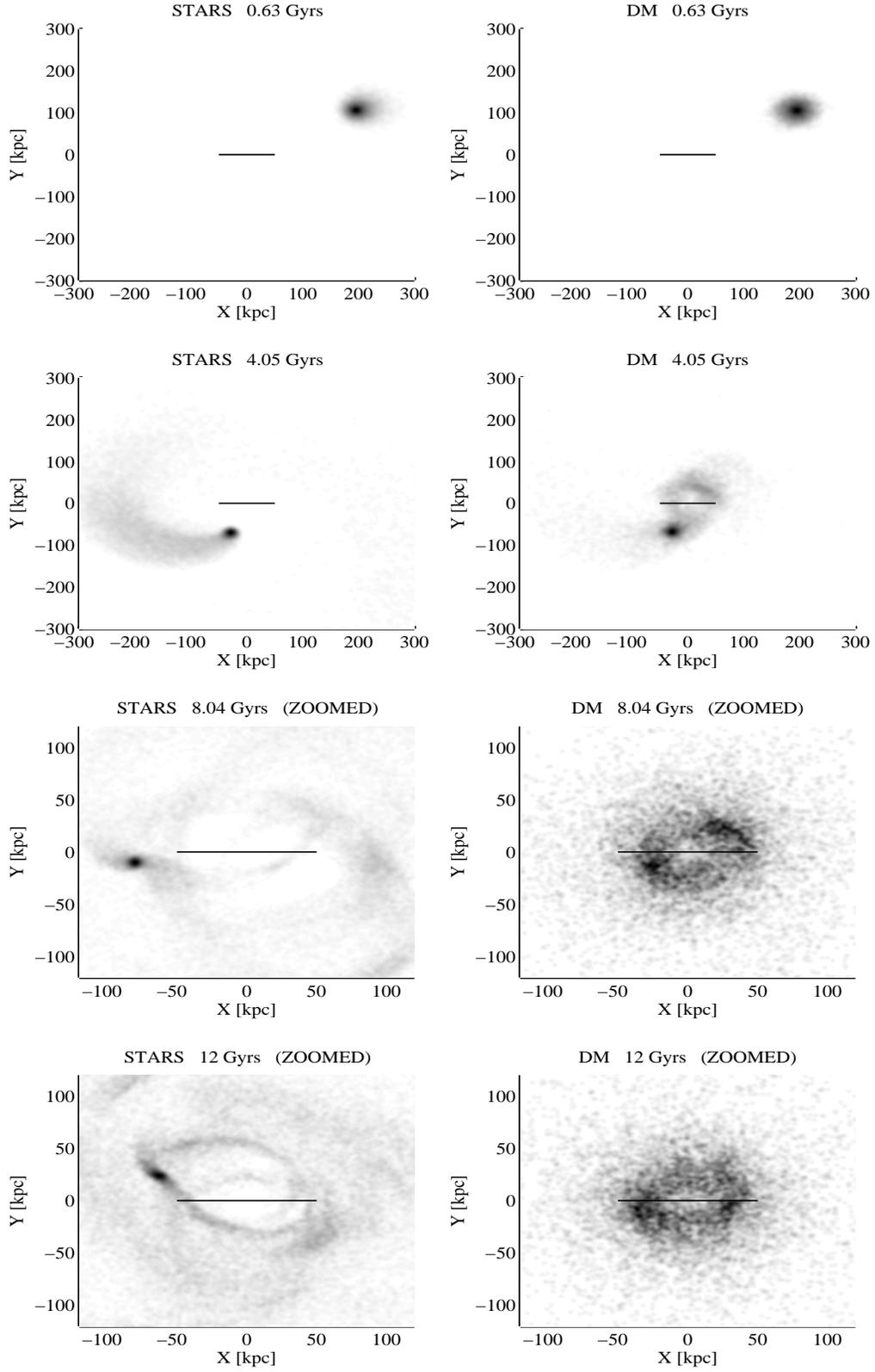, width=440pt, height=700pt}}
\caption{Mass distribution for stars and DM in the `heavy' Sgr simulation.}
\label{fig:fig17}
\end{figure*}

We have run a simulation of this `heavy' Sgr scenario,
including
a `live' MW halo to account for  dynamical friction. The results are illustrated in Fig.~\ref{fig:fig17}.
Here, the MW halo is modeled with an NFW profile of
concentration $15$, mass of $2 \times 10^{12} \ms$, and virial radius $250 \kpc$. The number of particles is
$2 \times 10^4$, and the smoothing $8 \kpc$. The Galactic disk and bulge have the same parameters as in the previous
simulations, however the bulge is not a static potential, but a single massive particle smoothed at $9 \kpc$ with a 
spline kernel, as all particles in Gadget2, and
the center of the stellar disk static potential follows the particle that represents the bulge. The initial
Sgr has a baryonic component of mass $4 \times 10^{10} \ms$ with a truncation radius of
$3.6 \kpc$ and is represented by $7 \times 10^4$ particles, smoothed at $55 \pc$. The DM component is five times
more massive, has a virial radius of $29 \kpc$ represented with $2 \times 10^4$ particles, and smoothed at
$850 \pc$. Initially, the Sgr is positioned in the stellar-disk plane at a GC distance equal to the Galactic
virial radius of $250 \kpc$, and its velocity is set to the baryonic circular velocity of $185 \kms$, perpendicular
to the stellar-disk plane.
        
At the initial position in this simulation the 
mass and concentration of the Sgr DM halo are large enough to prevent 
separation between stars and DM. This is important, because while the
baryons are bound to the DM core, the orbital dynamics is almost
independent of $\beta$, since changing $\beta$ is equivalent to a change in
the gravitational constant $G$, which in turn is equivalent to a
transformation in the time variable
$t \rightarrow t \times \sqrt{\beta+1}$.
Hence, the Sgr
begins to sink in the usual way, while experiencing dynamical friction
and loosing mass to the MW. Only when the Sgr is in an orbit similar to the one 
which is inferred directly from the observations
 (i.e. with an apocenter of $60 \kpc$),  the DM halo 
becomes small 
enough to allow full separation. Afterwards the Sgr develops symmetric tidal arms,
in a similar way to S1 and S2, and because it looses about 99\% of its mass,
it is similar to the Sgr today. This works also for larger turn-around
distances, since as mentioned above, the dynamics are independent of
$\beta$ as long as the DM and stellar cores are initially tightly bound.   
\bigskip

\section{CONCLUDING REMARKS}
We have addressed some of the issues in astronomy that are raised
by the postulate of a strong departure from the weak equivalence
principle by nonbaryonic dark matter. One must also consider the 
effect on structure formation  on larger scales. A preliminary analysis \cite{r38} 
of the effect of adding a strong LRSI to 
the $\Lambda$CDM cosmology indicates an improved 
fit of theory to observations of voids, which
are more completely emptied by the LRSI, and of giant galaxies, which
complete major mergers sooner.
Since the DM clusters more rapidly in the LRSI scenario, the baryons
on scales below $\rsc$ would be expected to lag behind the DM.
This effect could leave a fraction of the baryons out of galactic haloes
and help explain the missing baryon problem reviewed in \cite{r42}.
We plan to
present a more detailed and systematic analysis of these considerations 
using the numerical methods developed here with the further adjustment of the
numerical code to take account of the departure from an LRSI inverse
square law in Eq.~(\ref{eq:eq1}).

As we noted in the last section there are still other issues to consider. One is whether a satellite
with all the properties of Draco may naturally form in the strong LRSI
picture. Another is the excess satellite 
problem \cite{r35, r36, r37}.  We stress here that LRSIs do not necessarily imply DM free 
satellites. In fact, small but  high concentrated DM halos 
could easily trap the stellar components so that full segragation 
never happens. This means that LRSI may even help increase the M/L ratio
by yanking only the least bound stars.
An interesting quantitative test in a pure particle simulation 
of the formation of an $L_\ast$ galaxy would count satellites
that are entirely stripped of the particles that represent 
baryons, satellites that retain enough stellar particles to 
be visible, and satellites that fall in at low redshift and are
capable of producing symmetric and orphan stellar streams. 

We are considering a substantial departure from standard physics, but
it is in the little explored dark sector where our present physics is so
very simple as to seem possibly suspicious. Thus we feel that
explorations of more complicated dark sector physics, while
speculative, are worthwhile. Our studies of the observational 
tests of the speculative idea that there is  a long-range departure from 
the weak equivalence principle in the dark sector
have not proved to be discouraging so far. 

\section*{ Acknowledgment}
We have benefited from instructions on the properties of the Sgr system 
by Kathryn Johnston and Gerry Gilmore and from discussions of
numerical simulations of satellite galaxies with Mike Kesden and Glennys Farrar.
This research is supported in part in Israel by the
German-Israeli Foundation for Scientific Research and
Development, by the German-Israeli Foundation for
Research and Development, by the Asher Space Research
Fund, by the Israel Science Foundation
(grant No. 203/09), by the Asher Space Research
Institute, and by the Winnipeg Research Fund,
and in part in the USA by Princeton University.
We also acknowledge two anonymous referees whose comments helped improve this text.

\bibliography{keselman09}

\appendix
\section{Satellite construction}

\label{app:app1}
We follow the recipe in 
\cite{r17}, with minor adjustment to accommodate two components, 
stars embedded in
a more massive DM halo. In this method, the single particle 
distribution function $f$ in phase space is a function only of the 
dynamical constant $Q$, 
\begin{equation}
Q \equiv \varepsilon - \frac{L^2}{2{r_a}^2}\, ,
\label{eq:eq5}
\end{equation}
where $\varepsilon$ is the total binding energy per unit mass, $\bf{L}$
is the particle angular momentum vector per unit mass, and $r_a$ is
the scale of anisotropy, given as
\begin{equation}
1-\frac{{\sigma_\theta}^2}{{\sigma_r}^2} =
\frac{r^2}{r^2+{r_a}^2}.
\label{eq:eq6}
\end{equation}
The distribution function gives the number density,
\begin{equation}
\rho(r) = \int \! f(\varepsilon, L) \, d^3 v.
\label{eq:eq7}
\end{equation}
The inverse of this equation is \cite{r39}
\begin{equation}
f(Q) =
\frac{1}{\sqrt{8 \pi}} \left[ \int \! \frac{d^2\rho_Q}{d\psi^2} +
\frac{1}{\sqrt{Q}}
\left( \frac{d\rho_Q}{d_\psi} \right)_{\psi=0} \right]\; ,
\label{eq:eq8}
\end{equation}
where $\rho_Q \equiv \rho(r) (1+r^2/{r_a}^2)$ and
$\psi \equiv -\Phi-\Phi_s$ is the relative potential, including its
gravitational and scalar components.

We construct two distribution functions for the DM and the stars in
a satellite. The relative potential of the stars is
\begin{equation}
\Psi_b = \int_r^{\infty} \!
\frac{G}{r^2} \left[ \mbar(r)+\mdm(r) \right] \, dr\, ,
\label{eq:eq16}
\end{equation}
where $\mbar(r)$ and $\mdm(r)$ are the masses in stars 
and DM within radius $r$. The relative potential of the DM is 
\begin{equation}
\Psi_{dm} = \int_r^{\infty} \!
\frac{G}{r^2} \left[ \mbar(r)+(1+\beta)\mdm(r) \right] \, dr\, .
\label{eq:eq17}
\end{equation}

Convergence criteria \cite{r40} for numerical sampling  of the mass and velocity 
distribution functions require that the number of particles within the virial radius  satisfies
\begin{equation}
\nv > \frac{(2\cv)^4}{(\ln(1+\cv)-\cv/(1+\cv))^2}
\label{eq:eq18}
\end{equation}
for an NFW profile with concentration $\cv$, and the smoothing length $\epsilon$ is
\begin{equation}
\epsilon = 4 \frac {\rv}{\sqrt{\nv}}\; .
\label{eq:eq19}
\end{equation}
Under these conditions there are  enough particles that the
relaxation time is much longer
than the simulation dynamical time, and particle accelerations
 do not exceed the limit set by the smoothing
length that keeps the acceleration induced by two-particle interactions
weaker than the mean field halo acceleration.
 
Our numerical satellite construction code applied to a Hernquist potential
\cite{r12} yields a distribution function consistent with the analytic
solution to better than a part in $10^{5}$ except at the edges of the 
distribution.  We find similar consistency for the combined NFW and plummer profiles, where we numerically constructed the distribution functions from the analytic profiles and then numerically reconstructed from the distribution functions the density runs of the two components. 
\end{document}